# Surface enhanced Raman spectroscopy using 2D plasmons of InN nanostructures

Kishore K. Madapu, Sandip Dhara

Surface and Nanoscience Division, Indira Gandhi Centre for Atomic Research, Homi Bhabha

National Institute, Kalpakkam−603 102, India.

[*] Electronic Mail: madupu@igcar.gov.in; dhara@igcar.gov.in

**Abstract**: We explored the surface enhanced Raman scattering (SERS) activity of the InN nanostructures, possessing surface electron accumulation (SEA), using the Rhodamine 6G (R6G) molecules. SERS enhancement is observed for the InN nanostructures which possess surface electron accumulation (SEA). In case of high temperature grown InN samples, a peak is observed in the low wave number (THz region) of Raman spectra of InN nanostructures originating from excitation of the two dimensional (2D) plasmons of the SEA. The enhancement factor of four orders was calculated with the assumption of monolayer coverage of analyte molecule. SERS enhancement of InN nanostructures is attributed to the 2D plasmonic nature of InN nanostructures invoking SEA, rather than the contributions from 3D surface plasmon resonance (SPR) and chemical interaction. The role of 2D plasmon excitation in SERS enhancement is corroborated by the near-field light-matter interaction studies using near-field scanning optical microscopy.

**Introduction:**

Raman spectroscopy is an indispensable characterization technique in analytical chemistry as well as material science for its contribution to phase analysis. However, Raman spectroscopy is limited by its feeble scattering cross-section. For the non-resonant molecules the Raman scattering cross section is $10^{-32}$ cm$^2$/sr, where as in case of resonant molecules it is $10^{-27}$ cm$^2$/sr. The Raman scattering cross-section can be enhanced by means of resonance Raman scattering (RRS) using resonant excitation matching the band edge transition energy levels. In addition, Raman scattering cross-section is found to be enhanced enormously with the absorption of analyte molecules on the rough surfaces of noble metals. This phenomenon is called the surface enhanced Raman scattering (SERS) [1, 2]. The SERS enhancement factor can be as much as $10^{14}$- $10^{16}$, which makes it to realize a single molecule detection [3–5]. The enhancement of SERS is attributed to the combination of two phenomena comprising of, namely, electromagnetic mechanism (EM) and chemical mechanism (CM) [6–9]. Large contribution of the enhancement, however, is attributed to the EM effect and role of the CM is minimal with one to two orders of the value [10–12]. The EM mechanism exploits the surface plasmon resonance (SPR) phenomenon of metal nanoparticles and nano-protrusions [13]. The SPR prevails when the excited electrometric wave is coupled to the conduction band electrons of the metal nanostructures. As a consequence of the SPR, electric field near the surface region is enhanced by four orders, and the enhanced field is evanescent in nature. However, SPR is strongly dependent on the excitation frequency, which is closely matched with the resonance frequency of free electrons of the nanostructure. The resonance frequency of the coinage (Au, Ag and Cu) and transition (Pt, Pd) metal nanostructures fall in the visible region [14–16]. Consequently, most of the SERS enhancement measurements are carried on the metallic nanostructures, especially Ag and Au nanoparticles. However, CM relies on the charge transfer mechanism which results in the increase of molecular polarizability [15, 17−19].

Because of their exceptional enhancement, most of the SERS measurements are carried out on the metallic nanostructures [20, 21]. However, metal nanostructures suffer from their poor bio-compatibility and cost-effectiveness [22−24]. Recently, semiconductor nanostructures are being investigated as an alternative to metallic SERS substrates, which includes the metal oxide (MO) nanostructures, namely $TiO_2$, $ZnO$, $Fe_2O_3$, alongside compound semiconductor e.g., InAs, and elemental semiconductors of Si and Ge [15, 25−29]. Moreover, large dielectric constants of metal oxide nanoparticles ($Fe_3O_4$, $TiO_2$, $WO_3$, and $ZnO$) are also utilized to enhance the Raman scattering cross-section [30]. In addition, graphene is also reported for the SERS substrate showing enhancement of one order [12]. Because of the low carrier density in the semiconductors as compared to metals, the frequency of SPR is generally in the infrared region, which is far from the visible region [31]. As a result, SERS enhancement is expected to be very low in the process of EM. Thus, the SERS enhancement in the semiconductors can be attributed to CM through the charge transfer effect [28, 29, 32]. In the context of bio-compatibility, III-V nitrides, namely InN, GaN, and AlN are excellent candidates [33]. However, to the best of our knowledge till now there are no reports available as III-nitrides as SERS substrates. Especially, InN shows a unique property of possessing the surface electron accumulation (SEA) along its surface region [34−36]. The sheet carrier density in the SEA layer behaves like two dimensional electron gas (2DEG) [37−39]. The plasmonic nature of 2DEG is different from conventional 3D plasmons, where plasmon frequency varies with the carrier density. In contrast, resonance frequency depends on the in-plane wave vector ($k$) along with the areal carrier density in former case [38, 39].

In the present study, 2D plasmonic nature of the sheet carrier density in the SEA is explored for the SERS enhancement. The SERS activity of InN nanostructures is probed using the different laser excitations. In addition, the role of SEA induced 2D plasmons in the enhancement of Raman spectra is probed. The measurements were carried out on the standard SERS analyte of Rhodamine 6G (R6G).

**Experimental:**

InN nanostructures were grown via atmospheric chemical vapour deposition technique using the metallic In (99.999%) as the source and ultra high pure $NH_3$ (99.9999%) and the reactive gas. The $c$-$Al_2O_3$ was used as the substrate. Details of the growth procedure can be found elsewhere [40]. The growth of the InN nanostructures was carried out at three different temperatures of 580, 620, and 650 $^o$C for the present study. Morphology of nanostructures was studied using the field emission scanning electron microscopy (FESEM; AURIGA, Zeiss). The vibrational properties were studied using micro-Raman spectrometer (InVia, Renishaw) with 514.5 and 488 nm laser excitations. All the Raman spectra were collected in the backscattering geometry and were detected using the thermo-electric cooled CCD detector with the help of 1800 and 2400 gr·mm$^{-1}$ grating monochromatization for 514.5 nm and 488 nm excitations, respectively. The Raleigh line cut off frequency was 30 cm$^{-1}$. The laser power of 3 μW and the objective lens of 50X (NA ~ 0.75) were used in the present study. Near-field imaging was carried out using the near-field scanning optical microscopy (NSOM) with aperture probe (MultiView 4000, Nanonics, Israel). In NSOM measurements, 532 nm laser was excited through the 100 nm size aperture and near-field distance was controlled using atomic force microscopy (AFM) feedback mechanism.

The SERS substrate was prepared by a quite simple approach. The $Al_2O_3$ substrate with InN ($Al_2O_3$/InN) nanostructures grown on it was immersed in the $10^{-3}$ M R6G solution for a while (1 or 2 sec) and subsequently dried under the IR lamp. In literature, several methods were adopted for the measurement of SERS enhancement factors [11]. One of the simple approaches is to collect Raman spectrum in the presence and in the absence of SERS substrate with the same concentration of the analyte and experimental conditions. In the present study, we utilized the unintentionally made scratches on the sample as bare substrate. In other words, Raman spectrum was collected from the area of scratches on the sample where there were no nanostructures and was compared with the Raman spectrum collected from the area with InN nanostructures.

**Results and Discussion:**

The morphological features are shown for InN nanostructures grown at different temperatures such as 580 °C (Fig. 1a), 620 °C (Fig. 1b) and 650 °C (Fig. 1c). The FESEM micrographs revealed the fact that nanostructures were grown with the random size and shapes.

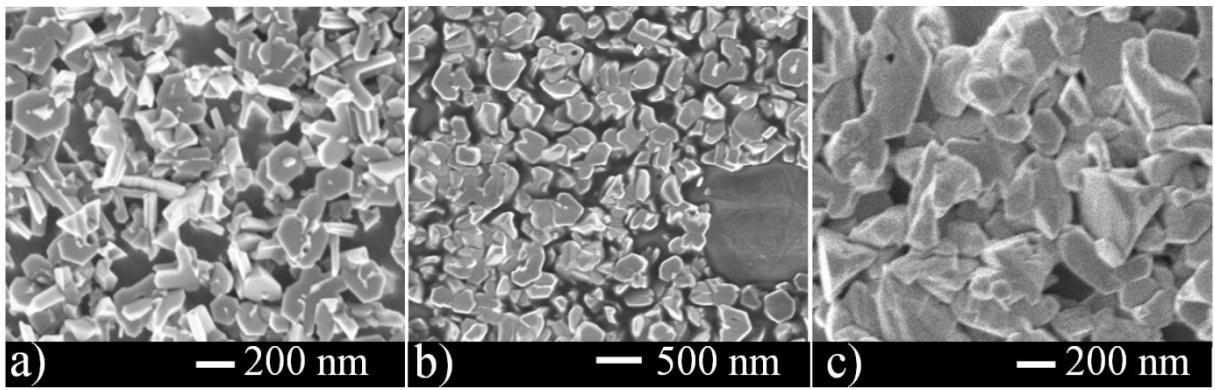

**Fig. 1** Morphology of InN nanostructures grown at different temperatures **a**) 580 °C **b**) 620 °C and **c**) 650 °C

However, the average size of the nanostructures was observed in the range of 100-200 nm. Raman spectroscopic analysis of these nanostructures is shown in Fig. 2 with the 514.5 nm excitation. The peaks observed in Raman spectra are readily assigned with symmetry allowed vibrational modes of the wurtzite-phase of InN belonging to the $p6_3mc$ space group which predicts six active Raman modes, $\Gamma = 2A_1 + 2E_1 + 2E_2$ [40, 41]. An asymmetric broadening of the $A_1$(LO) mode is observed in high temperature grown samples at 620 and 650 °C. The asymmetric broadening of the $A_1$(LO) mode is attributed to Fano interference of the free carrier density and the phonon mode through Frolich interaction [40,42]. The observation of Fano line shape explicitly reveals the fact that high temperature grown samples possess high carrier density as compared to that for the low temperature grown sample. Native defects in InN system, N vacancies, as well as the presence of SEA may be the reason for the increased carrier density in the high temperature grown samples [40]. In addition, a broad peak at the low frequency region is observed in high temperature grown samples only. Asymmetric broadening of the $A_1$(LO) mode and low frequency broad peak are observed for all samples grown at and above 620 °C. The low frequency peak can be attributed to the plasmon oscillations of InN nanostructures. However, the observed plasmon peak frequency at ~54-60 cm$^{-1}$ (~1.6-1.8 THz) may not be attributed to the bulk plasmons because 3D plasmon frequency of InN is calculated as the 1968 cm$^{-1}$ (discussed in forth coming section). In addition, observed peak may not be related to In clusters because of Mie resonance of In nanoparticles in InN matrix is observed in the energy range of 0.7-1 eV. Moreover, the interband transitions of most metals happen in the visible to UV region. The new peak in the low wave number region may be originated because of excitation of 2D plasmons in the form of surface plasmon polaritons (SPPs) [43–45]. Moreover, observed blue shift (Supporting information, Fig. S1) of the low frequency mode for samples grown at higher temperature further substantiates our assumption. The observed frequency is closely matched with reported 2D plasmon frequency in the range of THz for the SEA in InN [38, 39]. The condition for the excitation of the SPP in the layered structure is $k_{\parallel}d \ll 1$, where $k_{//}$ is the in-plane wave vector and $d$ is the thickness of the layered structure. In backscattering geometry, the in-plane wave vectors can be written as the $k_{\parallel} = (2\pi/\lambda)$ (sin$\theta_1$+sin$\theta_2$), where $\theta_1$ and $\theta_2$ are the angles between the incident beam and normal to the surface and scattered beam and normal to the surface, respectively. The schematic of the scattering geometry is shown in the inset of middle spectrum of Fig. 2. Because of the random alignment nanocrystals with respect to the incident laser, the minimum and maximum in-plane wave vector transfer occurs at angles of 0 and 90 degrees, respectively. Considering $d$ as the width of the SEA of InN nanostructures in the range of 4-10 nm, at incident and scattering angles of 45 degrees the $k_{\parallel}$ can be calculated as ≈ 2√2$\pi/\lambda$ which meet the condition [$k_{\parallel}d$ (0.07−0.17)≪ 1] for the SPP excitation [38]. The presence of In-In adatoms on the surfaces of InN is the physical origin of SEA [46]. The surface states of In-In adatoms exist above the conduction band minima (CBM) owing to the narrow band gap of InN. These surface states donate the electron to conduction band acquiring the positive charge. The electrons are accumulated near the surface to compensate the positive charge leading to the SEA. However, the presence of the In-In adatoms at the surface depends the on the growth conditions [46]. In the present case, owing to the low thermal stability of InN, the presence of the In-In adatoms can be expected in high temperature grown InN samples. Consequently, the SEA is present in the high temperature grown samples. As a result, the low-frequency feature is observed only in case of the high temperature grown samples. Thus, Raman spectroscopic analysis emphasizes that nanostructures grown at and above 620 $^o$C possess SEA.

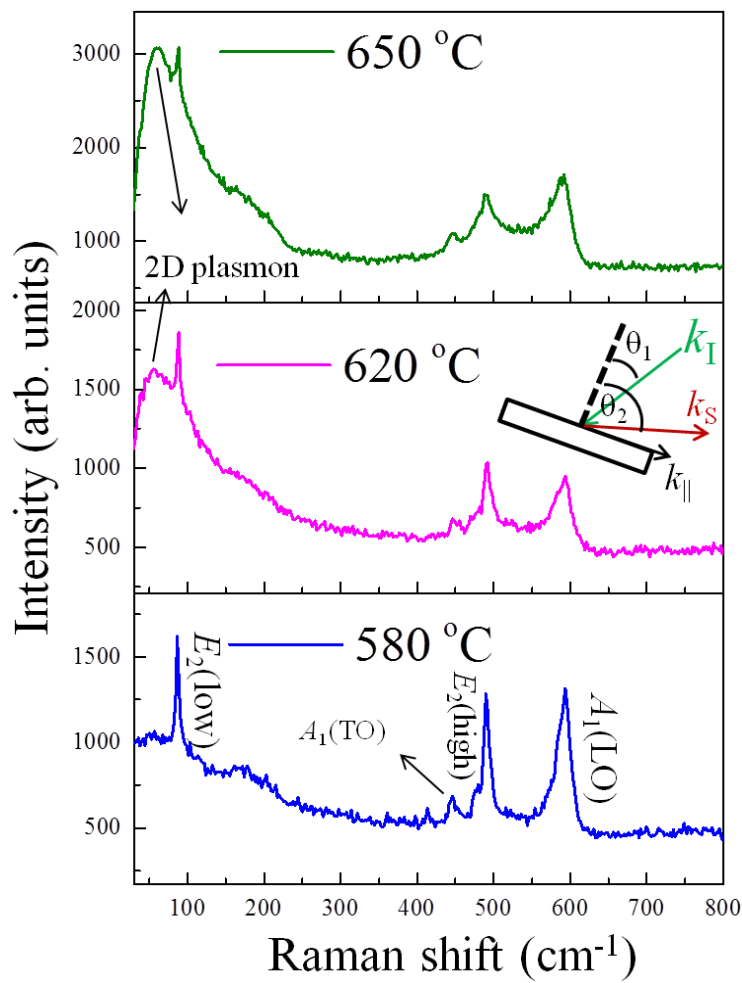

**Fig. 2** Raman spectra of InN nanostructures grown at 580, 620 and 650 °C. A low frequency peak, which is attributed to the 2D plasmon excitation, is evolved in high temperature grown samples. The schematic of the scattering geometry is shown in the inset of the middle spectrum.

Here, we explored the SERS activity of the InN nanostructures using the standard SERS analyte such as R6G. For measuring the SERS activity, two samples were selected which were grown at 580 and 650 $^o$C. Among these, high temperature grown sample possesses the SEA. Raman spectrum of the R6G molecules adsorbed on the InN nanostructures grown at 580 $^o$C is shown in Fig. 3a. The inset of Fig. 3a shows the optical image of the R6G molecules adsorbed on $Al_2O_3$/InN substrates. The circles in the inset figure represent the area for the collection of Raman spectra. The dashed circle represents the R6G on the bare $Al_2O_3$ substrate where the continuous circle represents the R6G on InN nanostructures. Raman spectra show the distinct peaks at frequencies of $\upsilon_1 = 614$ cm$^{-1}$, $\upsilon_2 = 774$ cm$^{-1}$, $\upsilon_3 = 1189$ cm$^{-1}$, $\upsilon_4 = 1309$ cm$^{-1}$, $\upsilon_5 = 1363$ cm$^{-1}$, $\upsilon_6 = 1512$ cm$^{-1}$, $\upsilon_7 = 1574$ cm$^{-1}$ and $\upsilon_8 = 1651$ cm$^{-1}$. The observed peaks were closely matched with reported values of R6G molecules. While other modes correspond to the aromatic symmetric stretching of the R6G molecule; $\upsilon_1$, $\upsilon_2$ and $\upsilon_3$ correspond to the C-C-C ring in–plane bending, out-of-plane bending of the H atoms of the xanthene skeleton, and C-C stretching vibrations, respectively [3, 47].

In the present study, Raman enhancement was calculated using the vibrational mode of $\upsilon_8 =$ 1651 cm$^{-1}$. Raman spectrum in case of the sample grown at 580 $^{\circ}$C without any SEA shows the same intensity along with overlap in the spectral features (Fig. 3a). It shows the negligible amount of enhancement of the intensity in the presence of the nanostructures grown at 580 $^{\circ}$C (Fig. 3a). In addition, the luminescence of the R6G molecules is collected in the presence and absence of the InN nanostructures (Fig. 3b), and no enhancement in the luminescence band is observed. A shoulder is observed (Fig. 3b) in the luminescence spectrum of R6G molecules on InN nanostructures $\sim$ 620 nm which is discussed subsequently.

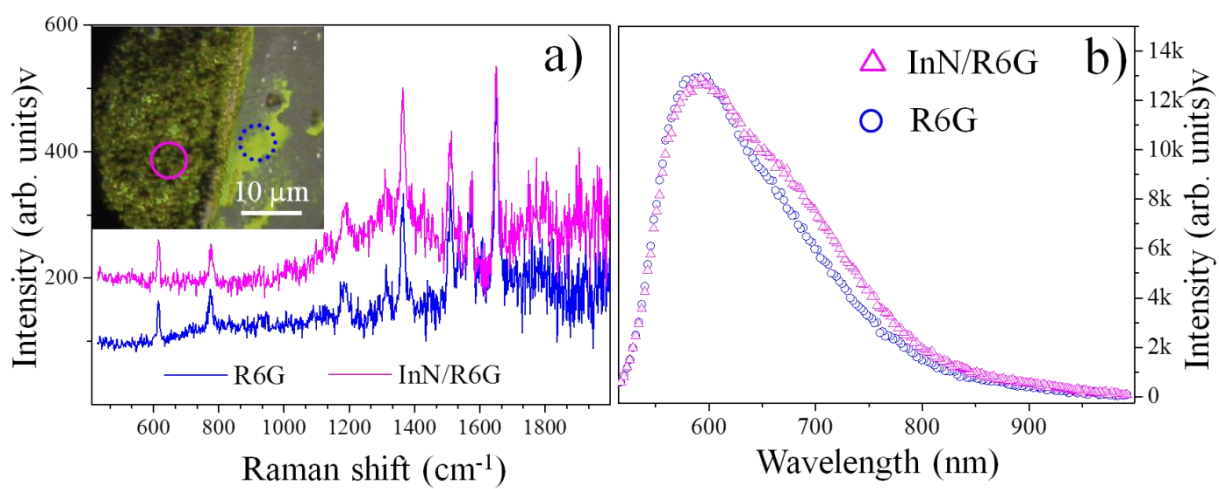

**Fig. 3** (**a**) SERS enhancement studies on InN nanostructures grown at 580 $^{\circ}$C. Inset figure shows the optical images of R6G adsorbed substrate with the presence (continuous circle; Al$_2$O$_3$/InN/R6G) and absence (dashed circle; Al$_2$O$_3$/R6G) of nanostructures where the Raman spectra were collected. (**b**) Luminescence band of 10$^{-3}$ M R6G in the presence and absence of the InN nanostructures.

The SERS enhancement of the InN nanostructures grown at 650 $^{\circ}$C is further carried out to study the effect of the SEA, as the nanostructures grown at and above 620 $^{\circ}$C are analyzed to possess prominent SEA (Fig. 2). The electrons in the SEA behave like absolute 2DEG. Raman spectra of the R6G molecules absorbed on the InN nanostructures, grown at 650 $^{\circ}$C, are shown in Fig. 4a. As compared the InN nanostructures grown at 580 °C, the spectra witness considerable amount of enhancement in the presence of InN nanostructures grown at 650 °C. The background correction was carried out using the cubic spline interpolation. The enhancement factor (EF) is calculated using the following equation,

$$EF = \frac{I_{SERS} N_{Normal}}{I_{Normal} N_{SERS}} \quad (1)$$

where $I_{SERS}$ is the intensity of a specific Raman mode of R6G; $I_{Normal}$ is the intensity of the same mode of R6G in the absence of SERS substrate; $N_{Normal}$ is number of R6G molecules in the excited volume in case of normal Raman analysis; and $N_{SERS}$ is the number of adsorbed R6G molecules for a single nanostructure with an average size.

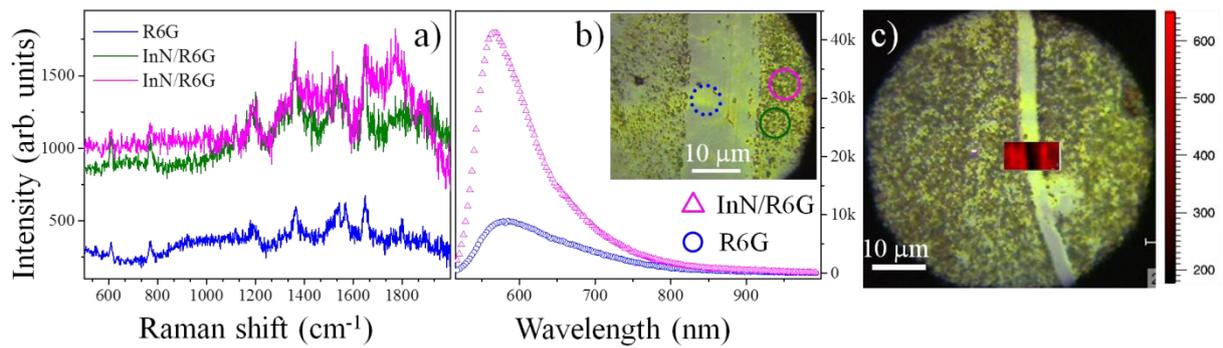

**Fig. 4** SERS and PL enhancement studies on InN nanostructures grown at 650 °C. **a**) Raman spectra collected in the presence (corresponding to the continuous circle) and absence (corresponding to the dashed circle) of InN nanostructures. **b**) Luminescence of R6G molecules in presence and absence of InN nanostructures. Inset of (b) shows the area of collection of Raman and PL spectra. **c**) Intensity mapping of Raman mode $\upsilon_8 = 1651$ cm$^{-1}$ showing the enhanced intensity in the presence of the InN nanostructures.

The $N_{Normal}$ can be calculated using the effective excitation volume, $V = \pi(D/2)^2 H$; where $D$ is the diameter of the beam size ($D \approx 0.84$ μm), and $H$ is the effective depth of the focus ($\approx 1$ μm). Subsequently, $N_{Normal}$ was calculated using the following equation,

$$N_{Normal} = (V\rho/M)N_A \quad (2)$$

where the $\rho$ and $M$ are the density (1.26 g/cm$^3$) and molecular weight (479.02 gm/mol) of the R6G; and $N_A$ is the Avogadro's number. The $N_{Normal}$ is calculated to be $\sim 3.5 \times 10^8$. The $N_{SERS}$ ($8.9 \times 10^4$) is evaluated using the average surface area of the nanostructures and molecular foot print of the R6G ($\sim 2.2$ nm$^2$) with the assumption of the self-assembled monolayer. However, $N_{SERS}$ depends on the surface coverage of the analyte molecule (mono or multilayer). Because of finite decay length (30 nm) of the enhanced field, the number of layers influenced by the enhanced field is more than one monolayer. The thickness of the monolayer of the R6G molecule is 1.2 nm [48]. Thus, the number of layers experiences the enhanced field is $\sim 25$ which is taken into account of $N_{SERS}$ calculation. Here, nanostructures are approximated as spherical particles with the size of 100 nm for the simplicity of calculation. The calculated EF is $\sim 1.4 \times 10^4$ for the Raman mode of $\upsilon_8 = 1651$ cm$^{-1}$. In addition, luminescence from the R6G molecules is collected in the presence and absence of InN nanostructures (Fig. 4c) showing enhancement of four times in the luminescence band. Similar to the sample grown at 580 $^o$C (Fig. 3b), a shoulder peak is also observed in the luminescence spectrum of the present sample with R6G (Fig. 4b). Rhodamine 6G (R6G) is one of the most frequently used dyes for applications in dye lasers and as a fluorescence tracer. The absorption and luminescence bands of R6G are dependent on the concentration of R6G molecules. This phenomenon is reported for aqueous solutions [49] and thin films [48]. The absorption and emission properties of R6G can be explained by the exciton model theory. According to the exciton theory, the dye molecule is considered as a point dipole. The dye molecules are configured as monomers (isolated molecules) and aggregate molecules (dimers and trimers). There are two basic dimer configurations such as perfectly sandwiched structure (H-type) and perfectly aligned structure (J-type). The former is characterized by intense absorption and forbidden emission because of dipole selection rules. In addition, the blue shift in the absorption edge is observed as compared to the monomer. The later configuration is fluorescent with a red shift in the absorption and luminescence bands as compared to the monomer. However, the dimer of R6G molecules can be configured with a distorted sandwich structure such as an oblique H-type dimer. This oblique dimer is fluorescent and possesses the characteristics of both H- and J-type dimer. The emission from the oblique dimer is red shifted as compared to monomer emission and depends on the angle between the transition moments of the dipole. In addition, excimer can also be formed by the association of excited molecule and unexcited molecule. Emission from the adsorbed molecules on the InN nanoparticles is strongly blue shifted as compared to the bulk crystallites (Fig. S2a). The observed shift corroborates to the fact that the adsorbed molecules are not aggregated. Moreover, the emission spectrum of adsorbed molecules is dominated by three main peaks centered at 556, 586, and 632 nm which correspond to monomer, dimer, and excimer, respectively (Fig. S2b). In contrast, the emission spectrum of bulk crystals is dominated by aggregate crystallites cantered around ~660 nm (Fig. S2c). These observations further confirm that adsorbed molecules are not aggregated. In addition, the emission spectrum collected from the scratch and nanoparticles area shows the negligible amount of shift (Fig. S2d). Raman imaging is carried out with the peak intensity of $\upsilon_8$ in an area which covers the nanostructures depleted region also. Raman intensity imaging (Fig. 4c) clearly shows the enhancement of peak intensity in the presence nanostructures. The SERS measurements are carried out in different areas of the substrates, which also show the similar kind of enhancement (Fig. S3). As the absorption edge of R6G (532 nm) is close to the 514.5 nm excitation, [29] the SERS studies are further carried out using the 488 nm laser excitation to avoid the luminescence back ground. In case of 488 nm excitation also, similar kind of enhancement in the Raman spectral intensity is clearly observed (Fig. 5).

Usually, SERS enhancement in the Raman spectra is contributed by both EM and CM. In case of semiconductors the plasmon frequency, $\omega_p$ is expressed as,

$$\omega_p = \sqrt{\left[\frac{4\pi n e^2}{m_e(\varepsilon_\infty + 2\varepsilon_m)}\right]} \qquad (3)$$

where $m_e$ is the effective electron mass ($\sim 0.07 m_0$), $\varepsilon_\infty$ is the high frequency dielectric constant (for InN $\varepsilon_\infty = 6.7$) [50] and $\varepsilon_m$ is the dielectric function of surrounding medium ($\varepsilon_{air} \sim 1$). Carrier density ($n \sim 2.53 \times 10^{19} \text{cm}^{-3}$) of InN nanostructures is estimated using the Burstein-Moss shift [40]. The plasmon frequency is calculated as 1968 cm$^{-1}$, which is far from excitation frequencies (514.5 nm = 19436 cm$^{-1}$; 488 nm = 20492 cm$^{-1}$) as well as the low frequency peak observed at $\sim 54$ cm$^{-1}$ for high temperature grown samples (Fig. 2).

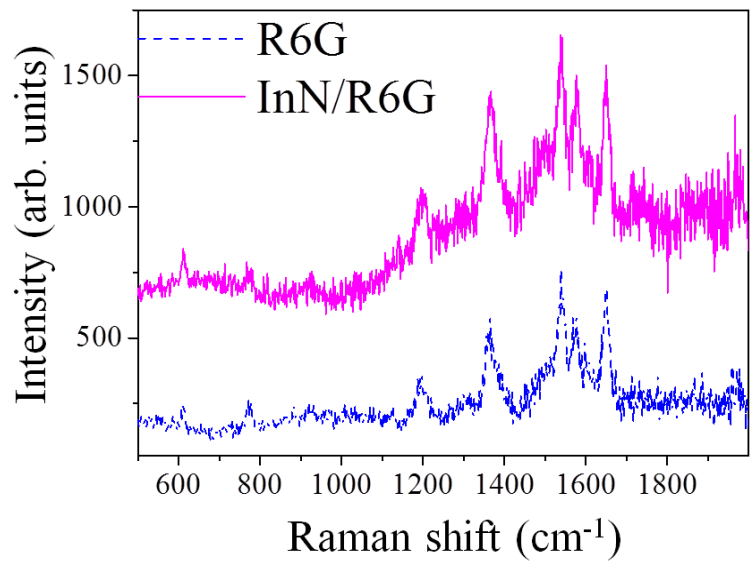

**Fig. 5** SERS studies of InN nanostructures grown at 650 $^o$C with excitation of 488 nm laser.

Thus, SPR cannot prevail in these nanostructures, and the role of EM can be ruled out in the enhancement of Raman spectral intensity. In general, SERS enhancement of the semiconductor nanostructures is attributed to the CM. In CM, the charge transfer takes place between the highest occupied molecular orbital (HOMO) levels of adsorbed molecules and lowest unoccupied molecular orbitals (LUMO) to the semiconductor energy levels. The charge transfer is feasible if the semiconductor valence band (VB) and conduction band (CB) levels closely match with the HOMO and LUMO levels of the adsorbed molecule [15]. In the present case, the R6G molecule has the LUMO and HOMO levels of −3.40 and −5.70 eV, respectively [29]. On the other hand, InN is calculated to have the VB of ~0 eV and the CB lies ~0.6-0.65 eV above the VB [51, 52]. These values indicated that charge transfer between R6G and InN may not be favourable because there is no close matching between R6G and InN energy levels. Moreover, in the present study, a variation in Raman spectral intensity enhancement was observed for the nanostructures grown at low (580 $^o$C) and high temperatures (650 $^o$C) (Figs. 3 and 4). It reveals that, there may be other reason than what it is usually attributed as CM for semiconducting materials. In order to understand the SERS enhancement, we studied the near-field light-matter interaction of these nanostructures using NSOM probe. NSOM imaging (Fig. 6) is carried out with the excitation of 532 nm laser using aperture probe of size 100 nm. Topography (Fig. 6a) and NSOM (Fig. 6b) image of nanostructures grown at 580 $^o$C show direct correlation between morphology and optical image of the sample, respectively. In addition, one can observe the high optical resolution image in NSOM because of the near-field imaging. Similarly, 620 $^o$C grown sample is also studied for topography (Fig. 6c) and NSOM image (Fig. 6d). In contrast to the near-field optical image of the sample grown at 580 $^o$C (Fig. 6b), high magnification NSOM image of 620 $^o$C grown sample (Fig. 6d) shows a clear enhancement of the near-field intensity around the nanostructures with the absorption of light. The height profile of one of the nanostructures (Fig. 6e) and its corresponding near-field intensity profile (Fig. 6f) show the strong absorption as well as enhancement of light intensity around the nanostructures. Similarly, 650 $^o$C grown sample is also further studied for topography (Fig. 6g) and NSOM image (Fig. 6h). In this case also enhancement in the near-field around the nanostructures is observed. Similar kind of field enhancement is observed in different NSOM images of nanostructures grown at 620 and 650 $^o$C (Fig. S4). Thus, near-field light-matter interaction analysis reveals that InN nanostructures with SEA show the enhancement for the light intensity in the vicinity of nanostructures.

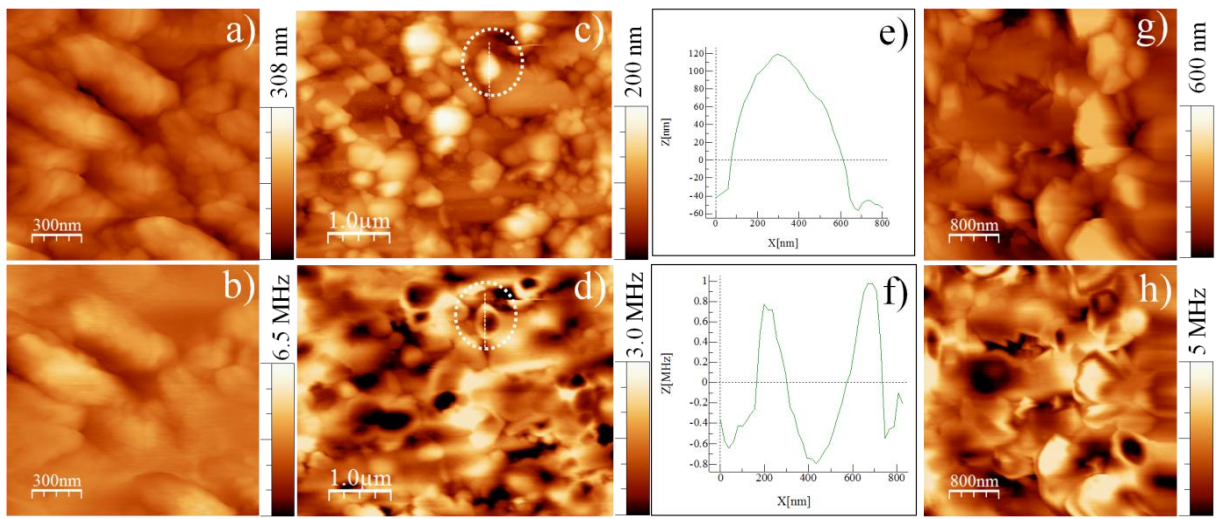

**Fig. 6** Near-field light-matter interaction of InN nanostructures grown at 580 $^o$C and 620 $^o$C and 650 $^o$C. (**a**) The topography of InN nanostructures grown at 580 $^o$C and (b) corresponding NSOM image. (**c**) The topography of InN nanostructures grown at 620 $^o$C and (**d**) corresponding NSOM image. (**e**) The height profile of a nanostructure indicated by a dotted circle in (**c**) and (**f**) its corresponding field distribution in NSOM image showing the enhanced field near the nanostructure surface. High magnification (**g**) topography and (**h**) NSOM image of InN nanostructures grown at 650 $^o$C.

The observed near-field enhancement of light intensity in the vicinity of nanostructures might be the reason for the SERS enhancement in case of nanostructures grown at 650 $^o$C.

However, observed field enhancement cannot be attributed to the SPR of bulk carrier density in the system because of the fact that 3D plasmon frequency is far from the excitation frequency. The enhancement of light intensity around the nanostructures may be originated because of the excitation of 2D plasmons of SEA as SPPs, where electrons behave like 2DEG. This argument is corroborated the by the observation of the low frequency peak in the Raman spectra of nanostructures which possesses the SEA. In the present study, Raman measurement conditions readily fulfill the condition of $k_{\parallel}d<<1$ for the excitation of 2D plasmons of InN SEA. The observed enhancement is attributed to 2D plasmon excitation and the EF is three orders higher than that for graphene sample [12]. The EF value is also found to be higher by one to two orders higher than that for some of the oxide based semiconducting substrates [15]. The observed EF value, however is substantially lower than the best values reported for 3D plasmons using noble metal nanostructures [3−5]. The SERS enhancement of metal nanoparticles is originated as confinement of light is smaller than the diffraction limit owing to localized surface plasmon resonance (LSPR)and subsequent field enhancement. Similarly, confinement of light also prevails in case of SPPs. However, confinement of light due to LSPR is expected to be very high because of the small size of metal nanoparticles ($a<<\lambda$, where 'a' is the size of the nanoparticles and $\lambda$ is the excitation wavelength). Moreover, the decay length of SPP is in the order of $\lambda/3−\lambda/5$. Thus, the confinement of light also has the same order and subsequently the field enhancement is expected to be low. In addition, a significant amount of total electric field energy of SPP mode resides inside the conductor. These are the factors which can be the reason for the observed modest enhancement in the Raman spectra of the 2D plasmon. The low frequency of SPP as well as a semiconducting absorbing medium also might be the reason for the moderate enhancement in the present case.

In summary, surface electron accumulation (SEA) dependent SERS activity of InN nanostructures is explored using the Rhodamine 6G molecules. The SEA of InN nanostructures is confirmed by the Raman spectroscopic analysis with the observation of low frequency 2D plasmon peak at ~54-60 cm$^{-1}$ (~1.6-1.8 THz). InN nanostructures with SEA show the considerable enhancement in the Raman spectral intensity. The enhancement factor is calculated to be ~ $1.4 \times 10^4$. Rather than the 3D surface plasmon resonance of noble metal nanostructures as well as the chemical mechanism of semiconducting substrates, the SERS enhancement is attributed to the excitation of the 2D plasmons of 2DEG corresponding to the SEA of InN. The observed moderate enhancement factor is attributed to the large decay length of SPP and subsequent low confinement of light in contrast to LSPR. Excitation of the 2D plasmons is confirmed using light-matter interaction studies in the near-field scanning optical microscopy. Results indicate that materials with the SEA, namely, InN and InAs can also be used as the semiconductor SERS substrates.

**Acknowledgement**

We would like to thank Avinash Patsha, Research Associate, Tel Aviv University, Israel, A. K. Sivadasan, Government Higher Secondary School, Pazhayannur-680587, Kerala, India and R. Pandian of SND, IGCAR for their technical inputs. We also thank S. R. Polaki for his help in the FESEM studies.

# References

1. J.R. Ferraro, *Introductory Raman spectroscopy* (Academic press, 2003)

2. E. Le Ru, P. Etchegoin, *Principles of Surface-Enhanced Raman Spectroscopy: and related plasmonic effects* (Elsevier, 2008)

3. S. Nie, S.R. Emory, Science **275**, 1102 (1997)

4. K. Kneipp, Y. Wang, H. Kneipp, L.T. Perelman, I. Itzkan, R.R. Dasari, M.S. Feld, Phys. Rev. Lett. **78**, 1667 (1997)

5. H. Xu, E.J. Bjerneld, M. Käll, L. Börjesson, Phys. Rev. Lett. **83**, 4357 (1999)

6. M. Moskovits, J. Raman Spetrosc. **36**, 485 (2005)

7. A. Hartschuh, Angew. Chem. Int. Ed. **47**, 8178 (2008).

8. M. Banik, A. Nag, P.Z. El-Khoury, A. Rodriguez Perez, N. Guarrotxena, G.C. Bazan, V.A. Apkarian, J. Phys. Chem. C **116**, 10415 (2012)

9. M. Moskovits, Phys. Chem. Chem. Phys. **15**, 5301 (2013)

10. B.N.J. Persson, K. Zhao, Z. Zhang, Phys. Rev. Lett. **96**, 207401 (2006)

11. E.C. Le Ru, E. Blackie, M. Meyer, P.G. Etchegoin, J. Phys. Chem. C **111**, 13794 (2007)

12. X. Ling, L. Xie, Y. Fang, H. Xu, H. Zhang, J. Kong, M.S. Dresselhaus, J. Zhang, Z. Liu, Nano Lett. **10**, 553 (2009)

13. K.A. Willets, R.P. Van Duyne, Annu. Rev. Phys. Chem. **58**, 267 (2007)

14. P.K. Jain, X. Huang, I. H. El-Sayed, M. A. El-Sayed, Plasmonics **2**, 107 (2007)

15. X. Wang, W. Shi, G. She, L. Mu, Phys. Chem. Chem. Phys. **14**, 5891 (2012)

16. N.C. Lindquist, P. Nagpal, K.M. McPeak, D.J. Norris, S.-H. Oh, Rep. Prog. Phys. **75**, 036501 (2012)

17. R. Esteban, A.G. Borisov, P. Nordlander, J. Aizpurua, Nat. Commun. **3**, 825 (2012)

18. M. Banik, P.Z. El-Khoury, A.Nag, A. Rodriguez-Perez, N. Guarrottxena, G.C. Bazan, V.A. Apkarian, ACS Nano **6**, 10343 (2012)

19. A.K. Sivadasan, A. Patsha, A.Maity, T.K. Chini, S. Dhara, J. Phys. Chem. C **121**, 26967 (2017)

20. S. Yampolsky, D.A. Fishman, S. Dey, E. Hulkko, M. Banik, E.O. Potma, V.A. Apkarian, Nat. Photonics **8**, 650 (2014)

21. K.T. Crampton, A. Zeytunyan, A.S. Fast, F.T. Ladani, A. Alfonso-Garcia, M. Banik, S. Yampolsky, D.A. Fishman, E.O. Potma, V.A. Apkarian, J. Phys. Chem. C **120(37)**, 20943 (2016)

22. S. Dey, M. Banik, E. Hulkko, K. Rodriguez, V.A. Apkarian, M. Galperin, A. Nitzan Phys. Rev. B **93**, 035411 (2016)

23. H.Y. He, S.T. Pi, Z.Q. Bai, M. Banik, V.A. Apkarian, R.Q. Wu, J. Phys. Chem. C **120(37)**, 20914 (2016)

24. J. Ju, W. Liu, C.M. Perlaki, K. Chen, C. Feng, Q. Liu, Sci. Rep. **7**, 6917 (2017)

25. L.G. Quagliano, J. Am. Chem. Soc. **126**, 7393 (2004).

26. Y. Wang, W. Ruan, J. Zhang, B. Yang, W. Xu, B. Zhao, J.R. Lombardi, J. Raman Spetrosc. **40**, 1072 (2009)

27. A. Musumeci, D. Gosztola, T. Schiller, N. M. Dimitrijevic, V. Mujica, D. Martin, T. Rajh, J. Am. Chem. Soc. **131**, 6040 (2009)

28. X. Wang, W. Shi, G. She, L. Mu, J. Am. Chem. Soc. **133**, 16518 (2011)

29. W. Ji, B. Zhao, Y. Ozaki, J. Raman Spetrosc. **47**, 51 (2016)

30. L. Li, T. Hutter, A.S. Finnemore, F.M. Huang, J.J. Baumberg, S.R. Elliott, U. Steiner, S. Mahajan, Nano. Lett. **12**, 4242 (2012)

31. J.M. Luther, P.K. Jain, T. Ewers, A.P. Alivisatos, Nat. Mater **10**, 361 (2011)

32. M. Mohammadpour, Z. Jamshidi, J. Phys. Chem. C **121**, 2858 (2017)

33. P. Sahoo, S. Suresh, S. Dhara, G. Saini, S. Rangarajan, A.K. Tyagi, Biosens. Bioelectron. **44**, 164 (2013)

34. I. Mahboob, T.D. Veal, C.F. McConville, H. Lu, W.J. Schaff, Phys. Rev. Lett. **92**, 036804 (2004)

35. P.D.C. King, T.D. Veal, C.F. McConville, F. Fuchs, J. Furthmüller, F. Bechstedt, P. Schley, R. Goldhahn, J. Schörmann, D.J. As, Appl. Phys. Lett. **91**, 092101 (2007)

36. W.M. Linhart, J. Chai, R.J. Morris, M. Dowsett, C.F. McConville, S. Durbin, T.D. Veal, Phys. Rev. Lett. **109**, 247605 (2012)

37. L. Colakerol, T.D. Veal, H.-K. Jeong, L. Plucinski, A. DeMasi, T. Learmonth, P.-A. Glans, S. Wang, Y. Zhang, L.F.J. Piper, Phys. Rev. Lett. **97**, 237601 (2006)

38. T.V. Shubina, A.V. Andrianov, A.O. Zakhar'in, V.N. Jmerik, I.P. Soshnikov, T.A. Komissarova, A.A. Usikova, P.S. Kop'ev, S.V. Ivanov, V.A. Shalygin, Appl. Phys. Lett. **96**, 183106 (2010)

39. T.V. Shubina, N.A. Gippius, V.A. Shalygin, A.V. Andrianov, S.V. Ivanov, Phys. Rev. B **83**, 165312 (2011)

40. K.K. Madapu, S.R. Polaki, S. Dhara, Phys. Chem. Chem. Phys. **18**, 18584 (2016)

41. V.Y. Davydov, V.V. Emtsev, I.N. Goncharuk, A.N. Smirnov, V.D. Petrikov, V.V. Mamutin, V.A. Vekshin, S.V. Ivanov, M.B. Smirnov, T. Inushima, Appl. Phys. Lett. **75**, 3297 (1999)

42. A.E. Miroshnichenko, S.Flach, Y. S. Kivshar, Rev. Mod. Phys. **82**, 2257 (2010)

43. S.J. Allen Jr, D.C. Tsui, R.A. Logan, Phys. Rev. Lett. **38**, 980 (1977)

44. D. Olego, A. Pinczuk, A.C. Gossard, W. Wiegmann, Phys. Rev. B **25**, 7867 (1982)

45. G. Fasol, N. Mestres, H.P. Hughes, A. Fischer, K. Ploog, Phys. Rev. Lett. **56**, 2517 (1986)

46.  C. G. Van de Walle, D. Segev, J. Appl. Phys. **101**, 081704 (2007)

47. P. Hildebrandt, M. Stockburger, J. Phys. Chem. **88**, 5935 (1984)

48. M. Chapman, M. Mullen, E. Novoa-Ortega, M. Alhasani, J.F. Elman, W.B. Euler, J. Phys. Chem. C **120**, 8289 (2016)

49. F.M. Zehentbauer, C. Moretto, R. Stephen, T. Thevar, J. R. Gilchrist, D. Pokrajac, K.L. Richard, J.Kiefer, Spectrochim. Acta, Part A **121**, 147 (2014)

50. K.K. Madapu, N.R.Ku, S. Dhara, C.P. Liu, A.K. Tyagi, J. Raman Spetrosc. **44**, 791 (2013)

51. I. Mahboob, T.D. Veal, L.F.J. Piper, C.F. McConville, H. Lu, W.J. Schaff, J. Furthmüller, F. Bechstedt, Phys. Rev. B **69**, 201307 (2004)

52. J. Wu, J. Appl. Phys. **106**, 5 (2009)

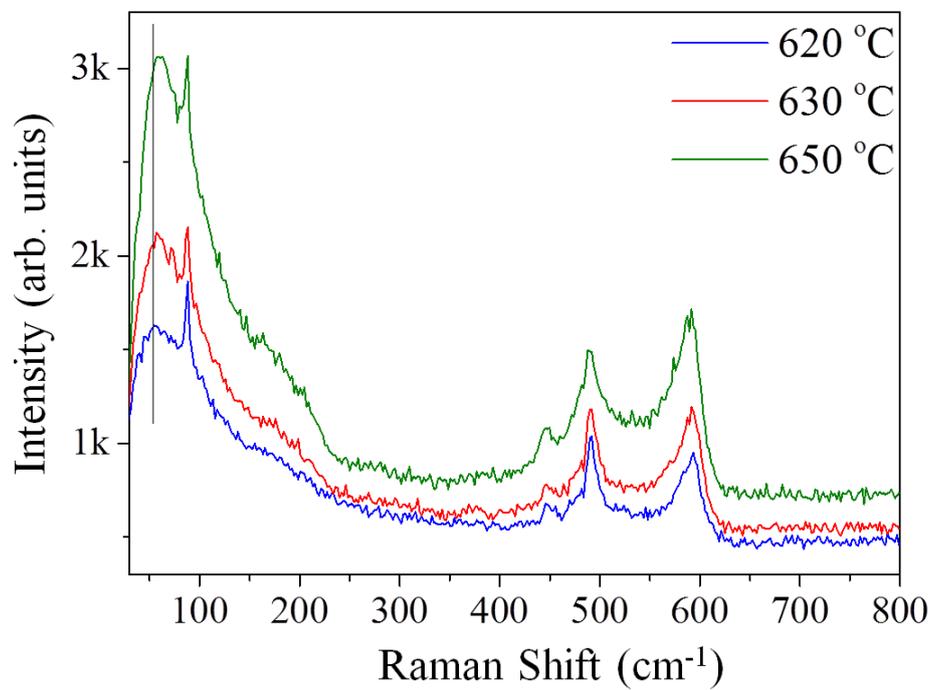

**Figure S1.** Variation of 2D plasmon peak frequency with increasing growth temperature.

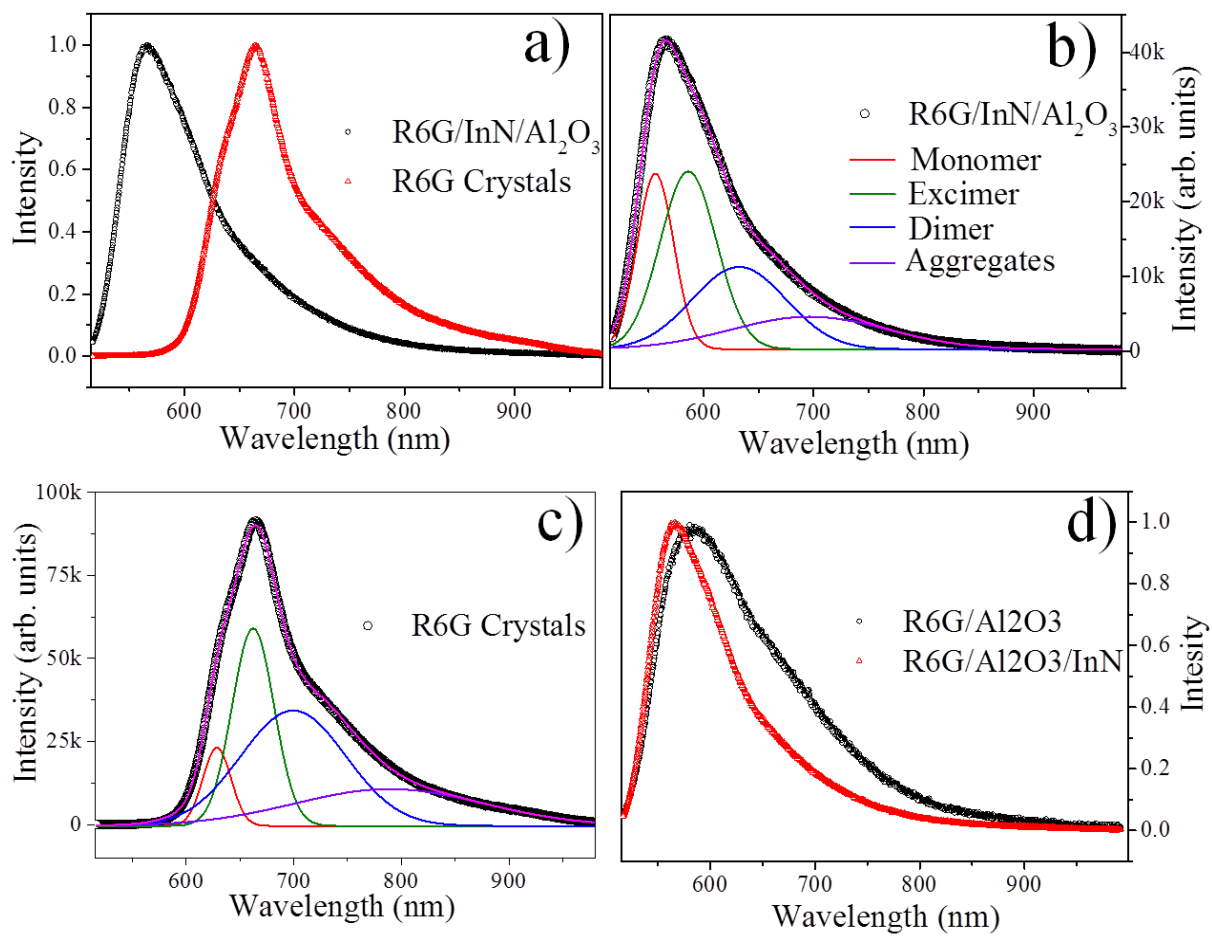

**Figure S2**. a) Normalized emission spectrum of the R6G adsorbed on the InN nanoparticles and R6G bulk crystals. b) and c) de-convoluted emission spectrum of adsorbed R6G on InN nanoparticles and R6G bulk crystals, respectively. The spectra were fitted with the Gaussian curves. d) Normalized emission spectrum of the R6G collected from scratch and nanoparticles area.

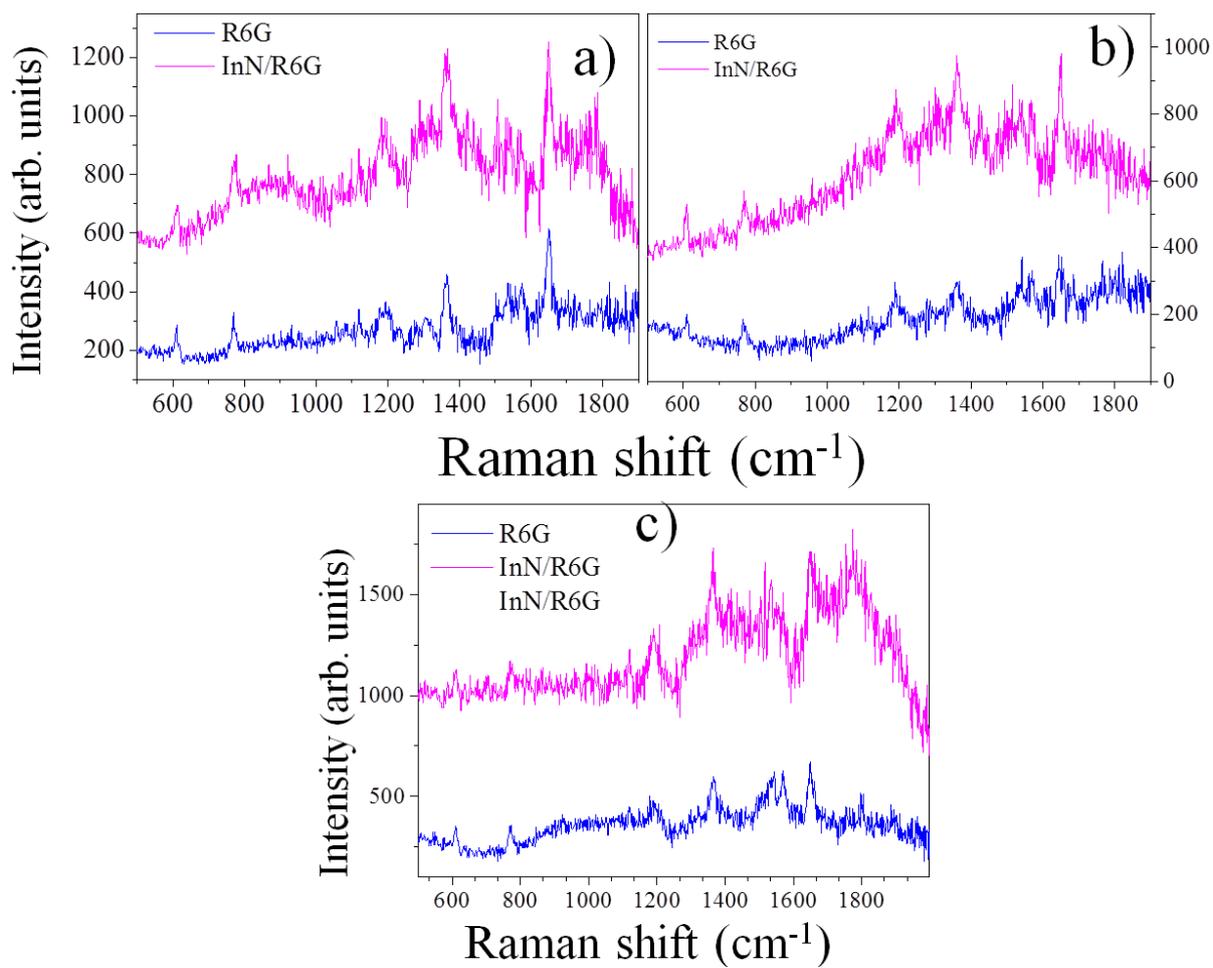

Figure S3. SERS enhancement studies on different positions of sample grown at 650 °C.

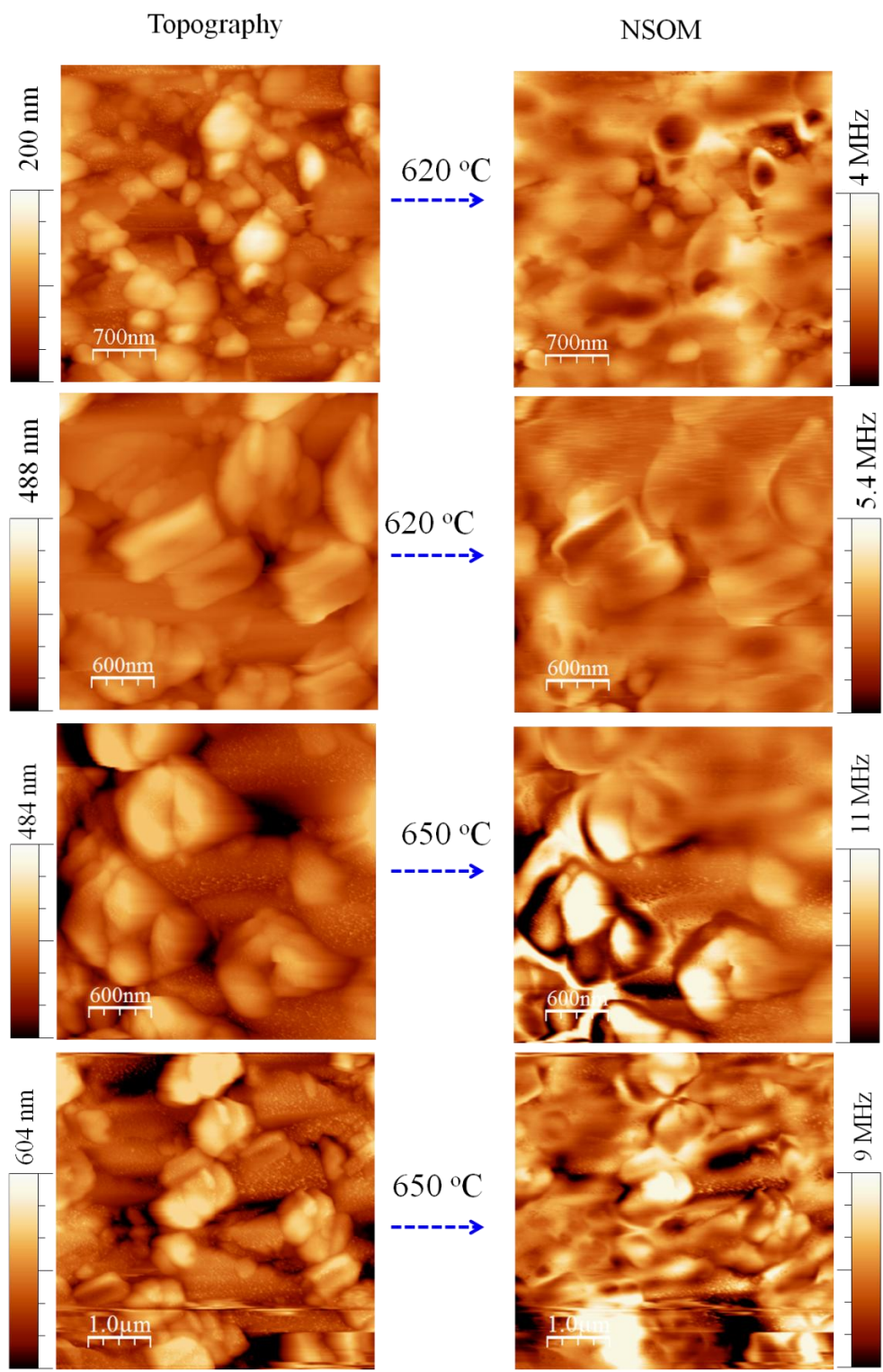

Figure S4. Near-field light-matter interactions in InN nanostructures grown at 620 and 650 °C possess surface electron accumulation. NSOM optical images show field enhancement close to the nanostructures.